\documentclass[prb,preprint,aps]{revtex4-1}
\usepackage{graphicx}
\usepackage{amsmath}
\usepackage{amssymb}
\usepackage{epsfig}
\usepackage{color}
\begin{document}


\title{Large diffusion lengths of excitons in perovskite and ${\it TiO_2}$  heterojunction}

\author{Zhyrair Gevorkian$^{1,2}$ ,Vladimir Gasparian$^{3}$ and Yurii Lozovik$^{4}$}
\affiliation{
$^{1}$ Yerevan Physics Institute,0036 Yerevan, Armenia. \\
$^{2}$ Institute of Radiophysics and Electronics,Ashtarak-2,0203,Armenia.\\
$^{3}$ California State University, Bakersfield  California 93311-1022,USA\\
$^{4}$ Institute of Spectroscopy, Academy of Sciences of Russia, 142092 Troitsk, Moscow Region. Russian Federation}




\begin{abstract}
Solar cells based on organometal halide perovskites have recently become very promising among other materials because of their cost-effective character and improvements in efficiency. Such performance is primarily associated with effective light absorption and large diffusion length of charge carriers. Our paper is devoted to the explanation of large diffusion lengths in these systems.
The transport mean free path of charged carriers in a perovskite/${\it TiO_2}$ heterojunction that is an important constituent of the solar cells  have been analyzed. Large transport length is explained by the planar diffusion of indirect excitons. Diffusion length of the coupled system increases by several orders compared to single carrier length due to the correlated character of the effective field acting on the exciton.
\end{abstract}


\maketitle
\section{Introduction}
Organometal Halide Perovskite ${\it CH_3NH_3PbI_{3-x}Cl_x}$ semiconductors have attracted great interest since their incorporation into the field of photovoltaics \cite{miyasaka09}. Exceptional performance of $>15$ percent  power conversion efficiency \cite{LJS13,Zhou14} is associated with the efficient light absorption and large electron-hole diffusion lengths \cite{Xing13, stranks13} (for a recent review see \cite{rev15}).
However, as it was shown in Ref. \cite{lin14}, an ordinary excitonic picture does not explain the existence of the large electron-hole diffusion lengths in perovskite semiconductors \cite{Xing13, stranks13,rev15} at room temperature. The main reason is that an estimate electron-hole coupling energy ($2meV$), based on
electro-optical properties of a perovskite solar cell, is small compared to the room thermal energy ($kT\sim 26meV$ ) (see also \cite{frost14}). Thus, a thermal recombination time is very small at room temperature and it is impossible to have a standard exciton transport in a single perovskite.

In this Letter we show that the experimental observed large transport lengths in perovskite solar cells \cite{Xing13, stranks13} that include an absorbing semiconductor as well as charge carrier transporting layers can be explained by involving the {\it spatially indirect} excitons (a coupled system of electron and hole that are trapped in different parallel planes, separated by some distance (see Fig.1)). Note, that although the binding energy of the indirect exciton in the single perovskite semiconductor is even smaller than the ordinary exciton's binding energy, however, it can be significantly larger for the geometry depicted in Fig. 1.  Indeed, if electron (hole) and hole (electron) get close to each other in space and at the same time belong to different layers, then significantly small static dielectric permittivity of the ${\it TiO_2}$ transporting layer ($\epsilon\sim 4$) and its thickness ($\sim 10 nm$) seems to be crucially important components for the formation of the indirect excitons. In this plausible scenario, the separation of electrons and holes in the direction perpendicular to the interface perovskite/${\it TiO_2}$ heterojunction (the growth
direction) reduces the wave function overlap, resulting in an increased exitons' lifetime. We will come back to this point later in more detail.

The realization of cold exciton gas in separated layers was proposed by one of the authors (Lozovik) with Yudson in Ref. \cite{lozyud76}. The indirect exciton model exhibits some similarities to a persistent current and the quasi-Josephson phenomena.  Because of this, it is not surprising that
during the past two to three decades many theoretical and experimental papers where devoted to  systems with indirect excitons (for a recent review see, e.g., Ref.\cite{sham15} and references therein). However, most of these systems were considered only at low helium temperatures (see, however Ref. \cite{butov}).
\section{Results}
\begin{figure}[ht]
\centering
\includegraphics[width=\linewidth]{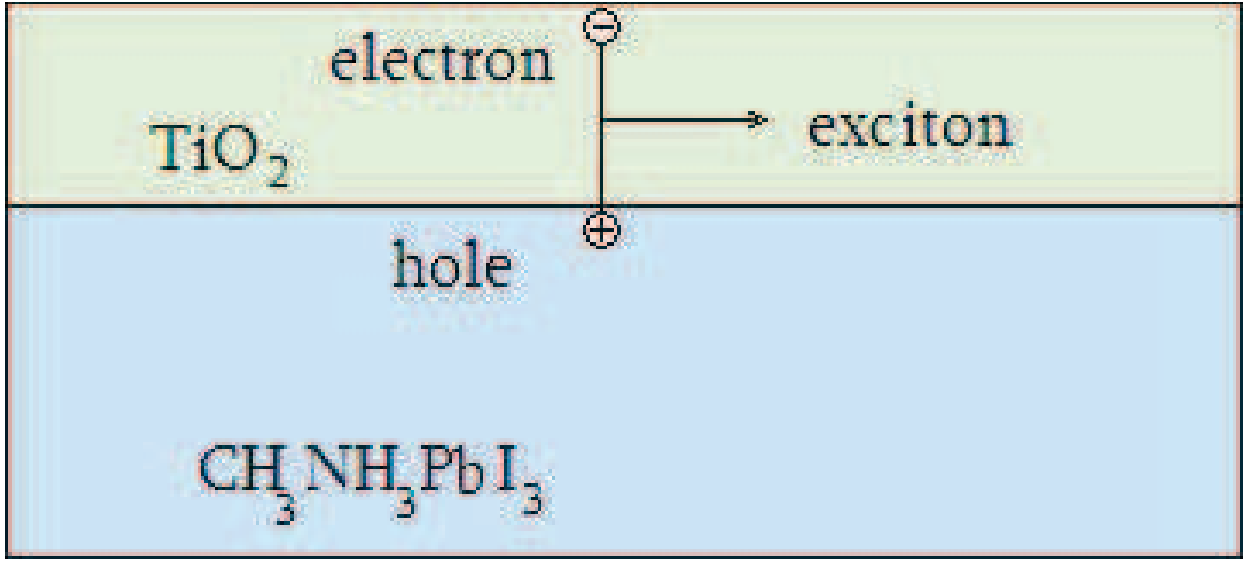}
\caption{Formation of indirect exciton from transport layer electron and perovskite hole.}
\label{fig:image2}
\end{figure}

Following Refs. \cite{lozyud76,lozruv97}, the Hamiltonian of an electron-hole pair in a perovskite/${\it TiO_2}$ heterojunction structure, presented in Fig. 1, can be recast into two-dimensional indirect exciton system, where electrons and holes are confined to different quantum wells and moved only parallel to the interface. This indirect exiton system can be described by the Hamiltonian
\begin{eqnarray}
\hat{H}=-\frac{\hbar^2}{2m_e}\nabla_{\bf r_e}^2-\frac{\hbar^2}{2m_h}\nabla_{\bf r_h}^2-\frac{e^2}{\varepsilon\sqrt{d^2+(\bf r_e-\bf r_h)^2}}
+\alpha_eV(\bf r_e)-\alpha_hV(\bf r_h),
\label{ham}
\end{eqnarray}
where $\bf r_e$ and $\bf r_h$ are two dimensional radius vectors of electron and hole, respectively. $m_e$ and $m_h$ are their effective masses , $d$ is the distance between electron and hole quantum wells, $\varepsilon$ is the static effective dielectric permittivity of the system and $V({\vec r}_e)$ ($V({\vec r}_h)$) is the random field of impurities seen by an electron (hole) along its movement path. $\alpha_e$ and $\alpha_h$ are constants that depend upon the band structure, impurity concentration and etc. It is worth noticing that although the Hamiltonian (\ref{ham}) does not contain any term describing the perovskite/${\it TiO_2}$ interface, its existence is implied. As it follows from the energy diagram (see, for example Ref. \cite{endiagr}) the electron transporting layer ${\it TiO_2}$ at the same time plays a role of barrier for holes due to the large difference between valence band energy values in perovskite and in ${\it TiO_2}$. This barrier hampers the penetration of the holes into ${\it TiO_2}$ forming a spatial separation of electrons and holes. According to this scenario, the electrons will remain in ${\it TiO_2}$ and holes in perovskite creating the indirect exiton states in two coupled quantum wells. The electron-hole pair moves parallel to the interface in these wells.

One of the main characteristic features of the perovskite semiconductors used in solar cell elements is that the mentioned impurities $V({\vec r}_e)$ and $V({\vec r}_h)$ are point-like potentials. They create shallow states either at the bottom (top of the valence band) or at the top (bottom of the conduction band) of the band gap \cite{mill14,yin14}.  Following this observation the field of impurities in the Hamiltonian (\ref{ham}) can
be modeled as a Gaussian distributed $\delta$-correlated random function
\begin{equation}
<V(\bf r)V(\bf \acute{r})>=g\delta(\bf r-\bf \acute{r}).
\label{corr}
\end{equation}
As was previously mentioned, formation of the exciton within a single perovskite film at room temperatures is impossible because of the large value of the static dielectric permittivity ($\sim 70$) and corresponding small value of exciton binding energy  (compared to $kT\sim 26meV$).

However, the formation of spatially indirect excitons is a feasible approach in the perovskite/${\it TiO_2}$ heterostructure, depicted in Fig. 1.  The transporting layer ${\it TiO_2}$ that is necessary in solar cells can serve simultaneously as a reservoir of electrons and as an opaque barrier for holes.  The bulk perovskite provides holes at the contact interface (see Fig.1), forming bounded couples with the electrons in ${\it TiO_2}$.  The latter has a relatively low static dielectric constant ($\sim 3-6$) (see Ref. \cite{Tio2}) and an estimated effective electron's mass equal to 10 of free electron masses (see, for example Ref.\cite{efmass}).  Due to the separation between electron and hole layers (typically 10nm) the average transport time of a bounded couple is at least $2,000$ times longer than the single charge carrier transport time. The latter leads to the large electron-hole diffusion lengths (see below for the details). This implies that one deals eventually with an indirect exciton.

It is convenient in Eq. (\ref{ham}) to use relative and center of mass coordinates instead of electron and hole coordinates: $\bf r=\bf r_e-\bf r_h$, ${\bf R}=({\bf r_e} m_e+{\bf r_h} m_h)/(m_e+m_h)$.  As a result the center of mass moves in an effectively random field  with a correlation function that is determined by the relative motion wave function, as is shown in Ref. \cite{gevloz85}.

The correlation function for the indirect exciton in the two dimensional case has a form \cite{berloz04}
\begin{equation}
B(R)=\frac{\alpha_h^2g}{2\pi^3\rho^2}\exp\bigg(-\frac{R^2}{\rho^2}\bigg),
\label{twocorr}
\end{equation}
where $\rho(d) = (8a)^{1/4}d^{3/4}$ is the radius of the indirect exciton and $a=\varepsilon \hbar^2/4\mu e^2$ is the two dimensional excitonic Bohr radius at $d\to 0$. The reduced mass $\mu$ of an electron and a hole system approximately equals to $m_h$, provided that $m_e\gg m_h$ and assuming that $\alpha_e\approx 0$, i.e. there are no defects in ${\it TiO_2}$. Note that the correlation radius of the effective field acting on the exciton is proportional to the electron and hole motion planes distance, which is essentially larger than the Bohr radius.
The next step, using the explicit form of the correlation function (\ref{twocorr}), is to estimate the transport time of an indirect exciton in the perovskite/${\it TiO_2}$ heterostructure.
We recall that the transport time of a particle moving in a correlated random field can be determined by (see, for example Ref.\cite{akkmont}).
\begin{equation}
\frac{1}{\tau}=2\pi\nu<B(\theta)(1-\cos\theta)>,
\label{trtime}
\end{equation}
where $<...>$ means averaging over the scattering angles. $B(\theta)$ is the Fourier transform of correlation function at the transferred momentum and $\nu$ is the density of states that in the weak scattering regime and in the two dimensional case will become particle and energy independent, i.e. $\nu=m/2\pi\hbar^2$.

Calculating explicitly the Fourier transform of the correlation function B(R), Eq. (\ref{twocorr}), we find
\begin{equation}
B(\theta)=\frac{\alpha_h^2 g}{16\pi^4}\exp\left[-\frac{p^2\rho^2(1-\cos\theta)}{8\hbar^2}\right],
\label{furtrans}
\end{equation}
where $p$ is the particle momentum. Substituting Eq. (\ref{furtrans}) into Eq. (\ref{trtime}) and taking the integral over the angle for the transport time in the limit $p\rho/\hbar>>1$, one gets
\begin{equation}
\tau\sim \frac{1}{\nu}\bigg(\frac{p\rho}{\hbar}\bigg)^3.
\label{trtimefin}
\end{equation}
This is our main result. Despite that this is only an approximate expression, it correctly predicts a tendency for an increasing transport time and the diffusion lengths in perovskite/${\it TiO_2}$ systems. Indeed, as seen from Eq. (\ref{trtimefin}), the time of the coupled exciton compared to a single particle's transport time ($\tau_0\sim 1/\nu$) is proportional to the large parameter $(p\rho/\hbar)^3$. Taking into account that the radius of the indirect exciton $\rho\sim d^{3/4}$, one obtains that the transport time is $\tau\sim d^{9/4}$ and thus strongly depends on ${\it TiO_2}$ film $d$ thickness. The latter plays a role of distance between the electron and hole quantum wells in a spatially indirect exciton scenario. The large factor $(p\rho/\hbar)^3$ will be shown to lead to: (i) a corresponding gain in exciton transport mean free path, which in the single particle and point-like scatterer case could be estimated  to be of the order of the average distance between scatterers, (ii) and a large diffusion length in perovskite solar cell systems \cite{stranks13,largediff14}.

Let us now estimate the gain factor taking into account more realistic parameters, based on Refs.\cite{stranks13,largediff14}. The exciton momentum $p$ is estimated as $p=\sqrt{2m_ekT}$, $m_e\approx 10m_0$ and $p\sim 2.7\times 10^{-25}kgms^{-1}$. The indirect exciton radius $\rho=(8ad^3)^{1/4}$ can be obtained taking $a=\varepsilon\hbar^2/4\mu e^2=a_B\varepsilon m_0/4\mu$, where $a_B=0.05nm$ is the Bohr radius and  the reduced mass $\mu\approx m_h\approx m_0$. Assuming $\varepsilon\sim 4$ and $d \approx 10nm$, one has $a\approx a_B$  and for a lower bound of the two dimensional exciton radius $\rho$ about $5nm$. However, $\rho$ will increase if the effective hole mass will be lighter than $m_0$ (see Ref.\cite{frost14}). Using the above mentioned numbers one obtains $(p\rho/\hbar)^3\sim 2000$. For $\varepsilon\sim 4$ and $d \approx 10nm$ we have for the indirect exciton binding energy $\sim e^2/\varepsilon d\sim 40meV$. This energy is larger than $kT\sim 26 meV$ and manifests itself such that at room temperature indirect excitons can be treated as metastable particles.

Finally, the same factor $(p\rho/\hbar)^3$ will lead to the corresponding increase in exciton's two dimensional diffusion  coefficient $D=v^2\tau/2$ and hence affects the indirect exciton diffusion length $L_D=\sqrt{Dt}$. Let us estimate the latter for  a characteristic carrier recombination time  $t=1ns$ (see Ref. \cite{felix}).
The transport mean free path $l=v\tau$ can be estimated as $l=l_0(p\rho/\hbar)^3$ ($l_0=n^{-1/3}$ is the average distance between two dimensional point-like scatterers and $n$ is the bulk concentration of impurities). Velocity is estimated as $v=\sqrt{2E/m}$, where kinetic energy $E\approx kT=26meV$ and $m\approx 10m_0$ is the effective mass of the exciton that is determined mainly by the effective electron mass in ${\it TiO_2}$.  Assuming a typical value for impurity concentration $n=10^{19}cm^{-3}$ (see for example, Ref. \cite{frost14}) for the exciton diffusion length we finally obtain $L_D\approx 10\mu m$. One can obtain even larger diffusion lengths by considering longer recombination times (see Ref. \cite{felix}) or lower impurity concentrations as can happen in single perovskite crystals \cite{bakr,dong}. Furthermore, the exciton-induced factor $(p\rho/\hbar)^3$ in transport time correspondingly will lead to increase of the charge carriers' mobilities, experimentally found in Ref.\cite{wehr}. Note that the two dimensional transport as a key of solar cell high performance was mentioned also in the recent paper \cite{rashk15}.
Here we consider the formation of the indirect exciton between an electron from transporting layer (${\it TiO_2}$) and a hole from perovskite. In an analogous manner one can consider the indirect exciton formed by the electron from perovskite and the hole from the transporting layer. Note, that the carrier diffusion lengths for perovskite without transporting layers are essentially small \cite{ellayer}, at least by factor $10^3$.

In summary, we have analyzed the transport mean free path and experimentally observed large diffusion length of charged carriers in perovskite/${\it TiO_2}$ heterojunctions. It has been argued that a non-trivial large diffusion lengths of charge carriers in perovskite/${\it TiO_2}$ system can be explained by involving the spatially indirect excitons (a coupled system of electron and hole that are trapped in different parallel planes, separated by some distance (see Fig.1)). The indirect excitons can be treated as metastable particles because their binding energy ($\sim 40meV$) is larger than thermal energy ($kT \sim 26meV$).
We have provided an analytical expression for the transport time, Eq. (\ref{trtimefin}), that correctly
predicts a tendency for an increasing transport time and the diffusion lengths $L_D$ ($\approx 10\mu m$) in perovskite/${\it TiO_2}$ systems.
\section{ Acknowledgments}
V.G. thanks J. Lofy for his careful reading of the manuscript. Yu.E.Lozovik was supported by project RNF 14-12-01080. Zh.G. is grateful to L.Matevosyan for helpful discussions.


\begin{thebibliography}{99}
\bibitem{miyasaka09} A.Kojima, K.Teshima, Y.Shirai, T.Miyasaka, {\it J.Am.Chem.Soc.} {\bf 131},6050-6051 (2009).
 \bibitem{LJS13} M.Liu, M.B.Johnston, H.J.Snaith, {\it Nature} {\bf 501},395-398,(2013).
 \bibitem{Zhou14}  H.Zhou, Qi Chen, Gang Li, Song Luo, Tze-bing Song, Hsin-Sheng Duan, Ziruo Hong, Jingbi You, Yongsheng Liu, Yang Yang,{\it Science }{\bf 345}, 542−546 (2014).
 \bibitem{Xing13} G.C.Xing, N.Mathews, S.Y Sun, S.S.Lim, Y.M.Lam, M.Gratzel, S.Mhaisalkar, T.C.Sum, {\it Science} {\bf 342}, 344-347,(2013).
\bibitem{stranks13} S.D.Stransks, G.E.Eperon, G.Granchini, C.Menelaou, M.J.P.Alcocer, T.Leijtens, L.M.Herz, A.Petrozza, H.J.Snaith, {\it Science} {\bf 342}, 341-344,(2013).
\bibitem{rev15} A.K.Chilvery, A.K.Batra, B.Yang, K.Xiao, P.Guggilla, M.D.Aggarwal, R.Surabhi, R.B.Lal, J.M.Currie, B.G.Penn, {\it J. Photon. Energy} {\bf 5(1)}, 057402 ,(2015).
\bibitem{lin14} Q.Lin, A.Armin, R.C.R.Nagiri, L.P.Burn and  P.Meredith, {\it Nature Photonics} {\bf 2014} 1 December online DOI: 10.1038/NPHOTON.2014.284.
\bibitem{frost14} J.M.Frost, K.T.Butler, F.Brivio, F.Hendon, C.H.M.Schilfgaarde and A. Walsh, {\it Nano Lett.} {\bf 14}, 2584–2590 (2014).
\bibitem{lozyud76} Yu.E.Lozovik and V.I.Yudson, {\it Zh. Eksp. Teor. Fiz.} {\bf 71}, 738-753,(1976).

\bibitem{sham15} Massimo Rontani and L. J. Sham, Coherent exciton transport in semiconductors,arXiv:1301.1726v2 [cond-mat.mes-hall], (2015).
\bibitem{butov} M.M.Fogler, L.V. Butov, K.S. Novoselov, {\it Nature Communications} {\bf 5} Article number: 4555 doi:10.1038/ncomms5555 (2014).
 \bibitem{lozruv97} Yu.E.Lozovik, A.M. Ruvinsky,{\it Physics Letters A} {\bf 227}, 271-284,(1997).
 \bibitem{endiagr} Seigo Ito, Soichiro Tanaka, Kyohei and Hitoshi Nishino, {\it J.Phys.Chem} {\bf 118},16995-17000,(2014).
 \bibitem{mill14} J.L.Miller, {\it Physics Today} {\bf 67}(5),13 (2014).
\bibitem{yin14} W.Yin, T.Shi, Y.Yan, {\it Appl.Phys.Lett.} {\bf 104}, 063903 (2014).
\bibitem{Tio2} T.Wang, J.He, J.Zhou, J.Tang, Y.Guo, X.Ding, S.Wu, J.Zhao,{\it Journal of Solid State Chemistry}, {\bf 183},2797–2804,(2010).
\bibitem{efmass} P.V.V.Jayaweera, A.G.U.Perera, K.Tennakone, {\it Inorganica Chimica Acta} {\bf 361}, 707-711,(2008);
R.Summitt, N.F.Borrelli, {\it J. Phys. Chem. Solids} {\bf 26},921, (1965);
 C.Kormann, D.W.Bahnemann, M.R.Hoffmann, {\it J. Phys. Chem.} {\bf 92},5196,(1988).
 \bibitem{gevloz85} Zh.S.Gevorkyan and Yu.E.Lozovik, {\it Sov.Phys.Solid State }{\bf 27},1079,(1985).
 \bibitem{berloz04} O.L.Berman, Yu.E.Lozovik, D.W.Snoke and R.D.Coalson, {\it Phys.Rev.B} {\bf 70},235310,(2004).
\bibitem{akkmont} E.Akkermans and G. Montambaux, {\it Mesoscopic Physics of Electrons and Photons},CAMBRIDGE UNIVERSITY PRESS (2007).
\bibitem{largediff14} S.Sun, T.Salim, N.Mathews, M. Duchamp, C.Boothroyd, G.Xing, T.C.Sum, and Y.M.Lam, {\it Energy Environ. Sci.}, {\bf 7},399.(2014).
\bibitem{felix} F.Deschler, M.Price, S.Pathak, L.E. Klintberg, D.D. Jarausch,
R.Higler,S. Hüttner,T.Leijtens, S.D.Stranks, H.J. Snaith, M. Atatüre,
R.T.Phillips, and R.H.Friend,{\it The Journal of Phys.Chem.Letters},{\bf 5},1421-1426,(2014).
\bibitem{bakr} D.Shi, V. Adinolfi, R. Comin, M. Yuan, E. Alarousu,
A.Buin, Y.Chen,S.Hoogland,A.Rothenberger,K.Katsiev, Y.Losovyj,X.Zhang, P.A.Dowben,
O.F.Mohammed, E.H.Sargent,O.M.Bakr,{\it Science}, {\bf 347},519-522,(2015).
\bibitem{dong} Q.Dong, Y.Fang,Y.Shao,P.Mulligan,J.Qiu, L.Cao,and J.Huang,{\it Science}  {\bf 347},  967-970 (2015).
\bibitem{wehr}C.Wehrenfennig,E.Giles, G.E. Eperon, M.B.Johnston , H.J. Snaith,
and L.M.Herz, {\it Adv. Mater.} {\bf 26}, 1584–1589,(2014).
\bibitem{rashk15}S.N.Rashkeev, F. El-Mellouhi, S.Kais and  F.H.Alharbi, {\it Scientific Reports} {\bf 5}: 11467, Doi 10.1038/srep 11467, (2015) .
\bibitem{ellayer} S.A.Bretschneider, J. Weickert, J.A. Dorman and L.Schmidt-Mendea,{\it APL MATERIALS} {\bf 2}, 040701 (2014).
\end{thebibliography}
\end{document}